\documentclass[prb,preprint,amsmath]{revtex4}

\usepackage{longtable}

\begin{document}

\title{Categorization of problems to assess and improve proficiency as teachers and learners}

\author{Chandralekha Singh}

\affiliation{Department of Physics and Astronomy, University of Pittsburgh, Pittsburgh, Pennsylvania 15260}


\begin{abstract}
We describe how graduate students categorize introductory mechanics
problems based on the similarity of their solutions. Graduate students were asked at the end of a teaching assistant training
class to categorize problems from
their own perspective and from the perspective of typical introductory
physics students whom they were teaching. We compare their categorizations with the
categorizations by introductory physics students and physics faculty
who categorized the same problems. The utility
of categorization as a tool for teaching assistant training and faculty development
workshops is discussed. 
\end{abstract}

\maketitle

\section{Introduction}

To help students learn effectively instructors
should become familiar with their students' level of expertise at the beginning
of a course. Instruction should be designed to build on
what students already know to ensure that they acquire the desired
expertise as determined by the goals of a course.\cite{bransford,fred2,fred3,maloney,sanjay,intuition,schwartz,posner}

Physics experts often take for granted that introductory students
will be able to distill the
underlying physics principles of a problem as readily as experts can. However, beginning
physics students are usually much more sensitive to the context and
surface features of a physics problem than experts. If an instructor
teaches the principle of conservation of angular momentum with the example
of a spinning skater and gives an examination problem requiring the
use of the same principle in the context of a collapsing neutron
star under its own gravitational force, students may wonder what
this astrophysics problem involving a neutron star
has to do with introductory mechanics. Without
appropriate guidance, the spinning skater problem may look nothing
like a neutron star problem to a beginning student even though both problems
can be solved using the same physics principle. The difference
between what instructors and students ``see'' in the skater
and neutron star problems is due to the fact that physics experts
view physical situations at a much more abstract level than beginning
students who often are sidetracked by context-dependent features.\cite{fred,fred4,larkin1,larkin2,schoenfeld,chi3,mestre,hardiman}

A crucial difference between the problem solving strategies used by
experts in physics and beginning students lies in the interplay between
how their knowledge is organized and how it is retrieved to solve problems.
Categorizing various problems based on similarity
of their solutions can be a useful tool for teaching and learning.\cite{chi3,mestre,hardiman}
In a classic study by Chi et al.\cite{chi3} a categorization task
was used to assess introductory physics students' level of expertise
in physics. Introductory physics students were
asked to group mechanics problems into categories based
on the similarity of their solutions. They were also asked to explain the reasons
for their groupings. Unlike experts who categorize them based on the
physical principles involved to solve them, introductory students
categorized problems involving inclined planes in one category and
pulleys in a separate category.\cite{chi3}

In Chi et al.'s out-of-classroom study, 24 problems
from introductory mechanics were given to eight introductory physics
student volunteers (novices) and eight physics graduate student volunteers
(experts).\cite{chi3} There were no differences in the number of categories
produced (approximately 8.5 categories by each group on average) and
four of the largest categories produced by each student from both
groups captured the majority of the problems (80\% for experts
and 74\% for novices). Immediately after the first categorization,
each student was asked to re-categorize the same problems. The second
categorization matched the first categorization very closely. It was concluded that both experts and novices were able to categorize
problems into groups that were meaningful to them.\cite{chi3}

Further analysis of the data in Ref.~\onlinecite{chi3} showed that experts
and novices group their problems in different categories based
on their knowledge associated with the categories. Physics graduate
students (experts) were able to distill physics principles applicable
in a situation and categorize the problems based on those principles.
In contrast, novices based their categorization on the problem's
literal features. For example, 75\%, 50\%, 50\%, and 38\%
of the novices had springs, inclined plane, kinetic energy, and pulleys
as one of their categories, respectively; 25\%
of the experts used springs as a category but inclined plane, kinetic
energy, and pulleys, were not chosen as categories by any of the experts. 

A categorization task can also be used as a tool to help students
learn effective problem-solving strategies and to organize their knowledge hierarchically,
because such tasks can guide students to focus on the similarity of
problems based on the underlying principles rather than
on the specific contexts. For example, introductory physics students
with different levels of expertise can be given categorization tasks
in small groups, and students can be asked to categorize problems and
discuss why different problems should be placed in the
same group without asking them to solve the problems explicitly. Then
there can be a class discussion about why some categorizations are
better than others, and students can be given a follow-up categorization
task to ensure individual accountability. One advantage
of such an activity is that it focuses on conceptual analysis and planning
stages of problem solving and discourages the plug and chug approach.
Without guidance, students often implement a problem
solution without thinking whether a particular principle is applicable.\cite{fred2}

In this paper we report about the results of our study on the nature
and level of understanding of physics graduate students about the initial physics
knowledge of introductory students. We asked graduate students at the end
of a course for teaching assistants 
to categorize problems based on the similarity of their solutions, both
from their own perspective and from the perspective of an introductory
physics students. We compared their categorizations with those
performed by physics professors and introductory physics students.

One surprising finding is the resistance of graduate students to categorizing
problems from a typical introductory physics student's perspective
with the claim that such a task is ``useless", ``impossible", and has ``no
bearing" on their teaching assistant (TA) duties. Based on our finding, we suggest that
inclusion of such tasks can improve the effectiveness of TA training
courses and faculty development workshops and help TAs and instructors
focus on issues related to teaching and learning.

\section{Graduate Students in a TA Training Course}

We will discuss the process and outcome of the categorization
of 25 introductory mechanics problems by 21 physics graduate students enrolled in a TA training course at the end of the course. Graduate students
first performed the categorizations from their own perspective and
later from the perspective of a typical introductory student. 
The goals of the study were to investigate the following issues:

\begin{itemize}
\item How do graduate students enrolled in a TA training course categorize
introductory physics problems from their own perspective? 
\item How do graduate students categorize the same problems from the
perspective of a typical introductory physics student? Do they have
an understanding of the differences between their physics knowledge
structure and those of the introductory physics students? 
\item How does the categorization by the graduate students from their own
perspective compare with the categorization by introductory physics
students and physics faculty from their own perspective? 
\item How do introductory physics students in an in-class study categorize
the introductory mechanics problems after instruction compared to
the eight introductory student volunteers studied in Ref.~\onlinecite{chi3}? Does
the ability to categorize introductory mechanics problems by introductory
physics students depend strongly on the nature and context of the questions
that are asked? 
\end{itemize}

The issues involved in a detailed comparison with Ref.~\onlinecite{chi3} will be
discussed elsewhere. All those who performed the categorization were
provided the following instructions given at the beginning of the questions:\cite{epaps} 

$\bullet$ {Your task is to group the 25 problems below based upon similarity of solution
into various groups on the sheet of paper provided.
Problems that you consider to be similar should be placed in the same group.
You can create as many groups as you wish. The grouping of problems should NOT be in terms of
``easy problems", ``medium difficulty problems" and ``difficult problems" but rather it should be
based upon the features and characteristics of the problems that make them similar.
A problem can be placed in more than one group created by you. Please provide a brief explanation
for why you placed a set of questions in a particular group. You need NOT solve any problems.}\\
$\bullet$ {Ignore the retarding effects of friction and air resistance unless otherwise stated.}\\

The sheet on which individuals were asked to perform the categorization
of problems had three columns. The first column asked them to use their own category name for each of their categories,
the second column asked them for a description of the category that
explains why those problems may be grouped together, and the third
column asked them to list the problem numbers for the questions that
should be placed in a category. Apart from these directions, students were not given any other hints about the category names
they should choose.

\section{Types of Questions and Rating of Categories}

We were unable to obtain the questions in Chi et al.'s study
except for a few that have been published. We therefore chose
our own mechanics questions on sub-topics
similar to those chosen in Ref.~\onlinecite{chi3}. The context of
the 25 mechanics problems varied and the topics included one- and
two-dimensional kinematics, dynamics, work-energy, and impulse-momentum.\cite{epaps}

Many questions related to work-energy and impulse-momentum concepts
were adapted from an earlier study~\cite{chandra} and many questions on kinematics
and dynamics were chosen from other earlier studies~\cite{singhvideo,edit} because the development
of these questions and their wording had gone through rigorous testing
by students and faculty members.
Some questions could be solved using one physics principle for example, conservation
of mechanical energy, Newton's second law, conservation of momentum.\cite{epaps} 
The first two columns of Table 1 show the question numbers and examples of primary categories 
in which each question can be placed (based upon the physics principle used to solve each question).
Questions 4, 5, 8, 24 and 25 are examples of problems that involve the use of two principles for different parts
of the problem.\cite{epaps} 
Questions 4, 8, and 24 below can be grouped together in one category because they require the use of conservation of mechanical
energy and momentum:  

\begin{itemize}

\item
(4) Two small spheres of putty, A and B, of equal mass, hang from the ceiling on massless strings of equal length.
Sphere A is raised to a height $h_0$ as shown below and released. It
collides with sphere B (which is initially at rest); they stick and swing together to a maximum height $h_f$.
Find the height $h_f$ in terms of $h_0$.\\

\item
(8) Your friend Dan, who is in a ski resort, competes with his twin brother Sam on who can glide higher with the snowboard.
Sam, whose mass is 60 kg, puts his 15 kg snowboard on a level section of the track, 5 meters from a slope (inclined plane).
Then, Sam takes a running start and jumps onto the stationary snowboard. Sam and the snowboard glide together till they come
to rest at a height of 1.8 m above the starting level.
What is the minimum speed at which Dan should run to glide higher than his brother to win the competition?
Dan has the same weight as Sam and his snowboard weighs the same as Sam's snowboard.

\item
(24) You are standing at the top of an incline with your skateboard.  After you
skate down the incline, you decide to ``abort", kicking the skateboard out in
front of you such that you remain stationary afterwards.  How fast is the
skateboard travelling with respect to the ground after you have kicked it?
Assume that your mass is 60 kg, the mass of the skateboard
is 10 kg, and the height of the incline is 10 cm. 

\end{itemize}

Questions 5 and 25 below can be grouped together because they can be solved using conservation of mechanical energy and Newton's second law~\cite{edit}:

\begin{itemize}

\item
(5) A family decides to create a tire swing in their backyard for their son Ryan. They tie a nylon rope to a branch that is located
16 m above the earth, and adjust it so that the tire swings 1 meter above the ground. To make the ride more exciting, they
construct a launch point that is 13 m above the ground, so that they don't have to push Ryan all the time. You are their neighbor,
and you are concerned that the ride might not be safe, so you calculate the maximum tension in the rope to see if it will hold.
Calculate the maximum tension in the rope, assuming that Ryan (mass 30 kg) starts from rest from his launch pad.
Is it greater than the maximum rated value of 2500 N?

\item
(25) A friend told a girl that he had heard that if you sit on a scale while riding a roller coaster,
the dial on the scale changes all the time. The girl decides to check the story and takes a bathroom scale to the amusement park.
There she receives an illustration (see below), depicting the riding track of a roller coaster car along with information on the track
(the illustration scale is not accurate). The operator of the ride informs her that the rail track is smooth, the mass of the car is 120 kg, and that the car sets in motion from a rest position at the height of 15 m. He adds that point B is at 5m height and
that close to point B the track is part of a circle with a radius of 30 m.
Before leaving the house, the girl stepped on the scale which indicated 55 kg
(the scale is designed to be used on earth and displays the mass of the object placed on it).
In the rollercoaster car the girl sits on the scale. According to your calculation, what will the scale show at point B?

\end{itemize}

Although we had an idea about which categories created by individuals should be considered good or
poor, we validated our assumptions with other experts. We randomly selected the categorizations performed 
by twenty introductory physics students and gave it to three physics faculty who had taught
introductory physics recently and asked them to decide whether each
of the categories created by individual students should be considered good,
moderate, or poor. We asked them to mark 
each row which had a category name created by a student and a description
of why it was the appropriate category for the questions that were
placed in that category. If a faculty member rated a category created by an introductory student as good,
we asked that he/she cross out the questions that did not belong to
that category. The agreement between
the ratings of different faculty members was better than 95\%.
We used their ratings as a guide to rate the categories created
by everybody as good, moderate, or poor. A category was considered ``good'' only if it was based
on the underlying physics principles. We typically
rated both conservation of energy or conservation
of mechanical energy as good categories. Kinetic
energy as a category name was considered a moderate
category if students did not explain that the questions placed in that
category can be solved using mechanical energy conservation or the
work energy theorem. We rated a category such as energy as good if students
explained the rationale for placing a problem in that category. If
a secondary category such as friction or tension was the only category
in which a problem was placed and the description of the
category did not explain the primary physics principles involved, it was considered
a moderate category.
Table 1 shows examples of the primary and secondary categories and one
commonly occurring poor/moderate category for each question given in the categorization task. 

More than one principle or concept
may be useful for solving a problem. The instruction for the categorizations told
students that they could place a problem in more than one category.
Because a given problem can be solved using more than one approach,
categorizations based on different methods of solution that are appropriate
was considered good (see Table~1). For 
some questions, conservation of mechanical energy may be more efficient,
but the questions can also be solved using one- or two-dimensional kinematics
for constant acceleration. In this paper, we will only discuss categories that
were rated good. If a graph shows
that 60\% of the questions were placed in a good category by a particular
group (introductory students, graduate students, or faculty), it means that the
other 40\% of the questions were placed in moderate or poor
categories.

For questions that required the use of two major principles, those
who categorized them in good categories either made a category which
included both principles such as the conservation of mechanical energy and
the conservation of momentum or placed such questions in
two categories created by them -- one corresponding to the conservation
of mechanical energy and the other corresponding to the conservation of momentum.
If such questions were placed only in one of the two categories,
it was not considered a good categorization.

\section{Categorization by graduate students from Their Own Perspective}

A histogram of the percentage of questions placed in good
categories (not moderate or poor) is given in Fig.~1. This figure compares the average
performance of 21 graduate students at the end of a TA training course when they
were asked to categorize questions from their own perspective with
7 physics faculty and 180 introductory students who were given the
same task. Although this categorization by the graduate students is not on par with
the categorization by physics faculty, the graduate students displayed a higher
level of expertise in introductory mechanics than the introductory
students and were more likely to group the questions based
on physical principles. We note that in Ref.~\onlinecite{chi3} the experts were graduate students
and not physics professors.\cite{chi3}

Physics professors (and
sometimes graduate students) pointed out multiple methods for solving a problem
and specified multiple categories for a particular problem more often than the introductory students. Introductory students mostly placed one question
in only one category. Professors (and sometimes
graduate students) created secondary categories in which they placed a problem that were more like the introductory
students' primary categories. For example, in the questions
involving tension in a rope or frictional force,~\cite{epaps} many faculty and
some graduate students created these secondary categories called tension or friction, but also placed those questions in
a primary category, based on a fundamental principle of
physics. Introductory physics
students were much more likely to place questions in inappropriate
categories than the faculty or graduate students, for example, placing a problem that was
based on the impulse-momentum theorem or conservation of momentum
in the conservation of energy category. For questions involving two
major physics principles, for example, question 4 related to the ballistic
pendulum, most faculty and some graduate students categorized it in both the conservation of mechanical energy
and conservation of momentum categories in contrast to the introductory students
who either categorized it as an energy problem or as a momentum problem.
The fact that introductory students only focused on one of the principles involved to solve question 4 is consistent with an 
earlier study in which students either noted that this problem can be solved using conservation
of mechanical energy or conservation of momentum but not both.\cite{chandra}

Many of the categories generated by the three groups were the same, but
there was a major difference in the fraction of questions that were placed
in good categories by each group. What introductory students chose as
their primary categories were often secondary categories created by the
faculty. Rarely were there secondary categories made by the faculty,
for example, apparent weight, that were not created by students. There were some categories such as ramps, and pulleys, that were
made by introductory physics students but not by physics faculty or
graduate students. The percentage of introductory students who selected ramps, pulleys or even springs as categories 
(based mainly upon the surface features of the problem rather
than based upon the physics principle required to solve the problem) is significantly less (less than 15\% for each of these
categories) than in the study of Ref.~\onlinecite{chi3}. This difference
could be due to the fact that ours was an in-class study with a large
number of students and the categorization task was given a few weeks
after instruction in all relevant concepts. In contrast, in Ref.~\onlinecite{chi3}
there were only eight student volunteers and they might not have taken introductory mechanics recently. Another reason
for the difference could be due to the difference in questions that were
given to students in the two studies. In our study introductory students
sometimes categorized questions 3, 6, 8, 12, 15, 17, 18,
22, 24, and 25 as ramp problems, 
questions 6 and 21 as spring problems (question 21 was
categorized as a spring problem by introductory students who associated
the bouncing of the rubber ball with a spring-like behavior) 
and question 17 as a pulley problem. 
The lower number of introductory
students making spring or pulley as a category in our study could be due to the
fact that there are fewer questions than in Ref.~\onlinecite{chi3} that
involve springs and pulleys. However, ramp was a much less popular category
for introductory students in our study than in Ref.~\onlinecite{chi3} in which 50\% of the students created this category and placed at least
one problem in that category (although may questions can potentially be categorized as ramp problems even in our study).

Some introductory physics students created the categories speed
and kinetic energy if the question asked them explicitly
to calculate those physical quantities. The explanations provided
by the students as to why a particular category name, for example, speed,
is most suitable for a particular problem were not adequate; they wrote
that they created this category because the question asked for the
speed. Graduate students were less likely than introductory students
to create such categories and were more likely to classify questions
based on physical principles, for example, conservation of mechanical energy
(or conservation of energy which was taken to be a good category with proper explanation)
or kinematics in one dimension. Even if a problem did not
explicitly ask for the ``work done'' by a force on an object,
faculty and graduate students were more likely to create and place such questions
which could be solved using work-energy theorem or conservation of
mechanical energy in categories related to these principles. This task was much more challenging for the introductory physics students
who had learned these concepts recently. For example, it was easy
to place question 3 in a category related to work because the question
asked students to find the work done on an object.~\cite{epaps} Placing question~7 in the work-energy category was more difficult because
students were asked to find the speed.~\cite{epaps}

Figures 2--5 show histograms for questions 15, 21, 23, and 24 respectively of some common categories created by different percentages 
of introductory students, graduate students, and physics faculty.
(The categorization by graduate students from a typical introductory physics student's
point of view will be discussed in Sec.~V.) As expected, physics faculty performed most expert-like categorization for each of the problem
based upon the physics principles required to solve it followed by graduate students. For question 21 (see Figure 3) all
faculty created an impulse category while only 5 out
of 7 faculty also categorized it as a question related to momentum.
For this question categorization by all faculty was considered good. On the other hand, some graduate students and introductory students placed
this problem only in energy or force categories that were not considered good.

\section{Categorization by graduate students from Introductory Students' Perspective}

After the graduate students had submitted their own categorizations, they were asked
to categorize the same questions from
the perspective of a typical introductory physics student. A majority
of the graduate students had not only served as TAs for recitations, grading, or
laboratories, but had also worked during
their office hours with students one-on-one and in the Physics Resource Room at the University
of Pittsburgh. The goal of this task was to assess whether
the graduate students were familiar with the level of expertise of the introductory
students whom they were teaching and whether they realized that most introductory
students do not necessarily see the same underlying principles
in the questions that they do.
The graduate students were told that they were not expected to remember how
they used to think 4--5 years ago when they were introductory students. We wanted them to think about their experience as TAs in
introductory physics courses while grouping the questions from an introductory
students' perspective. They were also asked to specify whether they were recitation
TAs, graders, or laboratory TAs that semester.

The categorization of questions from the perspective of an introductory
physics student met with widespread resistance.
Many graduate students noted that the task was useless
or meaningless and had no relevance to their TA duties.
Although we did not tape record the discussion with the graduate students, we took
notes immediately following the discussion. The graduate students often asserted
that it is not their job to ``get into their students' heads.''
Other graduate students stated that the task was ``impossible'' and ``cannot
be accomplished.'' They often noted that they did not
see the utility of understanding the perspective of the students.
Some graduate students explicitly noted that the task was ``silly'' because
it required them to be able to read their students' minds and had
no bearing on their TA duties. Not a single graduate student stated that they saw
merit in the task or said anything in favor of why the task may
be relevant for a TA training course. The discussions with graduate students also suggest that many of them believed
that effective teaching merely involves knowing the content well and
delivering it lucidly. Many of them had never thought about the importance
of knowing what their students think for teaching to be effective.

It is surprising that most graduate students enrolled in the TA training course
were so reluctant or opposed to attempting the categorization task
from a typical introductory student's perspective. This resistance
is intriguing especially because the graduate students were given the task at the
end of a TA training course and most of them were TAs for introductory
physics all term. It is true that it is very difficult for the TAs
(and instructors in general) to imagine themselves as novices.
However, it is possible for TAs
(and instructors) to familiarize themselves with students'
level of expertise by giving them pre-tests at the beginning of a
course, listening to them carefully, and by reading literature
about student difficulties, for example, as part of the TA training course.

After 15--20 minutes of discussion we made the task more concrete
and told graduate students that they could consider categorizing from the perspective
of a relative whom they knew well after he/she took only one introductory
mechanics course if that was the only exposure to the material they
had. We also told them that they had to make a good faith effort even
if they felt the task was meaningless or impossible. Figure~6 shows
the histogram of how the graduate students categorized questions from their own perspective
and from the perspective of a typical introductory student/relative
who has taken only one physics course. Figure~7 shows the histogram
of how the graduate students categorized questions from the perspective of a typical
introductory student/relative in comparison to the categorization by introductory students. Figure~6 shows that the graduate students recognized
that the introductory physics students do not understand physics as
well as graduate students and hence they re-categorized the questions in worse
categories when performing the categorization from the perspective
of a typical introductory physics student (also see Figs.~2--5).
However, if we look at questions placed in each category, for example, conservation
of momentum, there are sometimes significant differences between the
categorization by graduate students from an introductory students' perspective
and by introductory students from their own perspective. This implies that while graduate students may have
realized that a typical introductory student/relative who has taken only one physics course may not perform as well as a physics graduate
student on the categorization task, overall they were not able to anticipate the frequency with which introductory students
categorized each problem in the common less-expert-like categories.

\section{Discussion}

The reluctance of TAs to re-categorize the questions from introductory students' perspective 
raises the question of what should the graduate students learn in a TA training class. In a typical TA training class, a significant
amount of time is devoted to writing clearly on the blackboard, speaking
clearly and looking into students' eyes, and grading students' work
fairly. There is a lack of discussion about the fact that teaching
requires not only knowing the content but understanding how students
think and implementing strategies that are commensurate with students'
prior knowledge and expertise.

After the graduate students had completed both sets of categorization tasks, we
discussed the pedagogical aspects of perceiving and evaluating
the difficulty of the questions from the introductory students' perspective.
We discussed that pedagogical content knowledge, which is critical
for effective teaching, depends not only on the content knowledge
of the instructor, but also on the knowledge of what the students
are thinking. The discussions were useful and many students
explicitly noted that they had not pondered why accounting for
the level of expertise and thinking of their students was important for devising strategies to facilitate learning. Some graduate students noted that they will listen to the introductory
students and read their written responses more carefully in the future.

One graduate student noted that after this discussion he felt that, similar to the
difficulty of the introductory students in categorizing the introductory
physics questions, he has difficulty in categorizing questions in the
advanced courses he has been taking. He added that when he is assigned
homework/exam questions, for example, in the graduate level electricity and
magnetism course in which they were using the classic book by Jackson,~\cite{jackson}
he often does not know how the questions relate to the material discussed
in the class even when he carefully goes through his class notes. The student noted that
if he goes to his graduate course instructor for hints, the instructor
seems to have no difficulty making those connections to the homework.
The spontaneity of the instructor's connection to the lecture material and the
insights into those questions suggested to the student that the instructor
can categorize those graduate-level questions and explain the method
for solving them without much effort. This facility is due in part because
the instructor has already worked out the questions and hence they have become an exercise. Other graduate
students agreed with his comments saying they too had similar experiences
and found it difficult to figure out how the concepts learned in the
graduate courses were applicable to homework problems assigned in
the courses. These comments are consistent with the fact that a graduate
student may be an expert in the introductory physics material related
to electricity and magnetism but not necessarily an expert in the
material at the Jackson level course. 
Such difficulty is not surprising considering that a handful of fundamental
physics principles are applied in diverse contexts. Solving questions
with different contexts involves transferring relevant knowledge from
the context in which it was learned to new contexts. The mathematical
tools required to solve the questions in advanced problems may increase
the mental load while solving questions and make it more difficult
to discern the underlying physics principle involved.

\section{Summary}

We found that
graduate students perform better at categorizing introductory mechanics
questions than introductory students but not as well as physics faculty.
When asked to categorize questions from a typical introductory physics
student's perspective, graduate students were very reluctant and many
explicitly claimed that the task was useless. This
study raises important issues regarding the content of TA training
courses and faculty professional development workshops and the extent
to which these courses should allocate time to help participants learn
about pedagogical content knowledge in addition to the usual discussions
of logistical issues related to teaching. Asking the graduate students and faculty to categorize questions from the
perspective of students may be one way to draw instructor's attention to these important issues
in the TA training courses and faculty professional development workshops.

\begin{acknowledgments}

We are grateful to Jared Brascher for his help in data analysis.
We thank F.\ Reif, R.\ P.\ Devaty, P.\ Koehler and J.\ Levy for useful
discussions. We thank all the students and faculty who performed the
categorization task and an anonymous reviewer for very helpful comments
and advice. We thank NSF for award DUE-0442087.
\end{acknowledgments}

\newpage\section*{Table}

\begin{table}[h]
{\tiny \begin{tabular}{|l|p{8.5cm}|p{4cm}|p{3cm}|}
\hline

Question & Examples of Primary Categories & Examples of Secondary Categories & Poor/Moderate Categories \\
\hline
1& (a) momentum conservation or (b) completely inelastic collision & & speed\\
\hline
2& (a) mechanical energy conservation or (b) 1D kinematics && speed\\
\hline
3& work by conservative force/definition of work& &ramp\\
\hline
4& mechanical energy conservation and momentum conservation & & only energy or momentum \\
\hline
5& mechanical energy conservation and Newton's second law & centripetal acceleration, circular motion/tension & only tension or only force\\
\hline
6 &mechanical energy conservation & & only spring\\
\hline
7 & work-energy theorem/definition of work or Newton's second law/1D kinematics & relation between kinetic energy & speed \\
\hline
8&momentum conservation or completely inelastic collision and mechanical energy conservation & & only energy or momentum\\
\hline
9&2D kinematics & &cliff\\
\hline
10& Newton's second law &circular motion/friction &only friction\\
\hline
11& linear momentum conservation or completely inelastic collision & &speed\\
\hline
12&mechanical energy conservation and work-energy theorem/definition of work & friction & only friction\\
\hline
13&Newton's second law &Newton's third law &force\\
\hline
14& 2D kinematics & &force/cliff\\
\hline
15&mechanical energy conservation & &speed\\
\hline
16&mechanical energy conservation & &speed\\
&or 2D kinematics& &\\
\hline
17&Newton's second law &Newton's third law/tension &only tension\\
\hline
18&mechanical energy conservation or 2D kinematics & & speed\\
\hline
19& Impulse-momentum theorem & & force\\
\hline
20& mechanical energy conservation& & speed\\
& or 2D kinematics& &\\
\hline
21&impulse-momentum theorem & & force\\
\hline
22& 2D kinematics & &ramp\\
\hline
23&Newton's second law/1D kinematics or &&force\\
&Work-energy theorem/definition of work&kinematic variables &\\
\hline
24&mechanical energy conservation or momentum conservation or completely inelastic collision & &speed\\
\hline
25&mechanical energy conservation and Newton's second law & centripetal acceleration, circular motion/normal force &ramp/force\\
\hline
\end{tabular}
}
\caption{\tiny \label{table1} Examples of the primary and secondary categories and one commonly occurring poor/moderate category for
each of the 25 questions}
\end{table}

\newpage\section*{Figure Captions}

\begin{figure}[h!]
\begin{centering}
\end{centering}
\caption{Histogram of introductory physics students, graduate
students, and physics faculty who categorized various percentages
of the 25 problems in ``good'' categories when asked to
categorize them based on similarity of solution from their own point
of view (SelfPoV). Good categories were determined in consultation
with three faculty members as discussed in Sec~III. 
Physics faculty performed best in the categorization task followed by graduate students and then introductory physics students.
}
\end{figure}

\begin{figure}[h!]
\begin{centering}
\end{centering}
\caption{Histogram of introductory physics students, graduate
students, and physics faculty who categorized various percentages
of question 15 in different categories when asked to group
them based on similarity of solution from their own point of view. The categorization by graduate students from a typical introductory
physics student's point of view is also shown.
Physics faculty performed best in categorization from their own point of view followed by graduate students and then introductory physics students.
When graduate students re-categorized problems from a typical introductory physics students' point of view, they grouped them in worse categories
(closer to the categories made by introductory students from their own point of view) than when they categorized from their own perspective.
Some category names have been abbreviated, e.g., the category ``energy" includes conservation of energy or conservation of mechanical energy.
}
\end{figure}

\begin{figure}[h!]
\begin{centering}
\end{centering}
\caption{Histogram of introductory physics students, graduate
students, and physics faculty who categorized various percentages
of question 21 in different categories when asked to categorize
them based on similarity of solution from their own point of view. Categorization by graduate students from a typical introductory
physics student's point of view is also shown. }
\end{figure}

\begin{figure}[h!]
\begin{centering}
\end{centering}
\caption{Histogram of introductory physics students, graduate
students, and physics faculty who categorized various percentages
of problem 23 in different categories when asked to categorize
them based on similarity of solution from their own point of view. Categorization by graduate students from a typical introductory
physics student's point of view is also shown. }
\end{figure}

\begin{figure}[h!]
\begin{centering}
\end{centering}
\caption{Histogram of introductory physics students, graduate
students, and physics faculty who categorized various percentages
of problem 24 in different categories when asked to categorize
them based on similarity of solution from their own point of view
(Self PoV). Categorization by graduate students from a typical introductory
physics student's point of view is also shown.}
\end{figure}

\begin{figure}[h!]
\begin{centering}
\end{centering}
\caption{Histogram of percentages of graduate students who categorized various
percentages of the 25 problems in ``good'' categories when
asked to categorize them based on similarity of solution from their
own point of view and when asked to categorize from the
perspective of a typical introductory physics students (As noted in
the text, the task was made more concrete later by asking graduate students to consider
a relative who had taken only one mechanics course.).}
\end{figure}

\begin{figure}[h!]
\begin{centering}
\end{centering}
\caption{Histogram of introductory physics students and graduate
students who categorized various percentages of the 25 problems in
good categories when asked to categorize them based
on similarity of solution from their own point of view 
and when asked to categorize from the perspective of a typical introductory
physics student respectively.}
\end{figure}

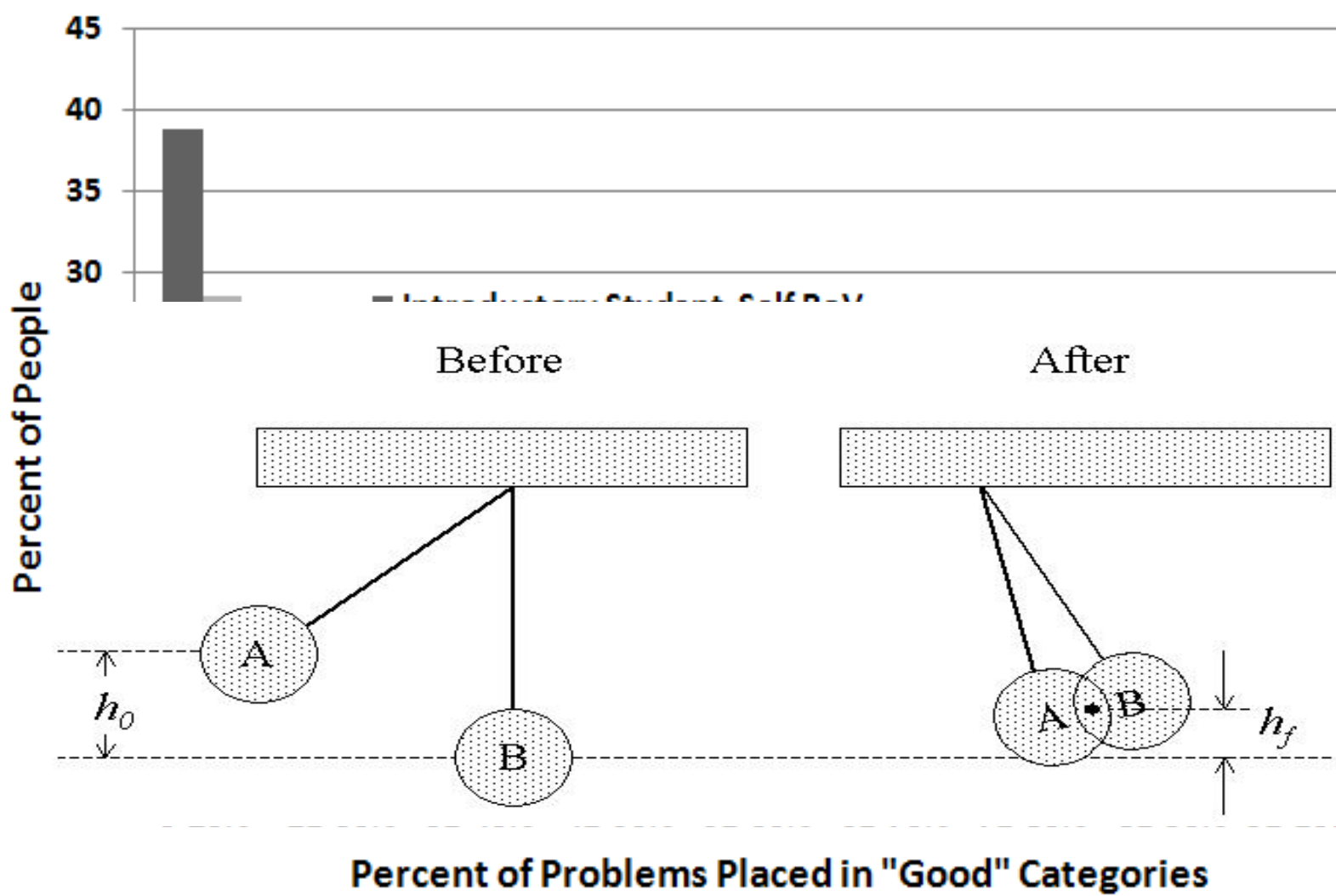
\end{document}